\documentclass[twocolumn,11pt, a4,superscriptaddress,notitlepage,preprintnumbers,aps,longbibliography]{scrartcl}
\usepackage{graphicx}
\usepackage{typearea}
\usepackage{amsmath}
\usepackage{subcaption}
\usepackage[noblocks]{authblk}
\usepackage{letltxmacro,xpatch}
\usepackage[colorlinks=true,
allcolors=blue,linktocpage=true]{hyperref}
\usepackage[switch]{lineno} \usepackage{listings}
\usepackage{xcolor}

\definecolor{codegreen}{rgb}{0,0.6,0}
\definecolor{codegray}{rgb}{0.5,0.5,0.5}
\definecolor{codepurple}{rgb}{0.58,0,0.82}
\definecolor{backcolour}{rgb}{0.95,0.95,0.92}

\lstdefinestyle{mystyle}{
    backgroundcolor=\color{backcolour},   
    commentstyle=\color{codegreen},
    keywordstyle=\color{blue},
    numberstyle=\tiny\color{codegray},
    stringstyle=\color{codepurple},
    basicstyle=\ttfamily\footnotesize,
    breakatwhitespace=false,         
    breaklines=true,                 
    captionpos=b,                    
    keepspaces=true,                 
    numbers=left,                    
    numbersep=5pt,                  
    showspaces=false,                
    showstringspaces=false,
    showtabs=false,                  
    tabsize=2
}

\typearea{18}
\title{The generative quantum eigensolver (GQE) and its application for ground state search}
\date{}

\usepackage{soul,color}

\newcommand{\tr}[0]{{\textrm{Tr}}}

\newcommand{\rhoinit}[0]{\rho_{\textrm{0}}}

\newcommand{\Mark}[0]{\footnote{These authors contributed equally.}}
\newcommand{\CoAuthorMark}[0]{\footnotemark[\arabic{footnote}]}

\DeclareMathOperator*{\argmin}{arg\,min}

\newcommand{\chemUofT}{1}
\newcommand{\AIST}{2}
\newcommand{\keio}{3}
\newcommand{\csUofT}{4}
\newcommand{\vecInst}{5}
\newcommand{\Harvard}{6}
\newcommand{\SJ}{7}
\newcommand{\IMS}{8}
\newcommand{\NVIDIA}{9}
\newcommand{\chemEngUofT}{10}
\newcommand{\matSciUofT}{11}
\newcommand{\CIFAR}{12}
\newcommand{\MCC}{13}
\author[\chemUofT,\AIST,\keio,\NVIDIA]{Kouhei Nakaji}
\author[\chemUofT,\csUofT]{Lasse Bjørn Kristensen}
\author[\keio]{Ryota Kemmoku}
\author[\chemUofT]{Jorge A. Campos-Gonzalez-Angulo \Mark}
\author[\chemUofT,\csUofT]{Mohammad Ghazi Vakili \protect\CoAuthorMark}
\author[\csUofT,\vecInst]{Haozhe Huang \protect\CoAuthorMark}
\author[\chemUofT,\csUofT]{Mohsen Bagherimehrab \Mark}
\author[\Harvard,\SJ]{Christoph Gorgulla \protect\CoAuthorMark}
\author[\csUofT,\IMS]{FuTe Wong}
\author[\NVIDIA]{Alex McCaskey}
\author[\NVIDIA]{Jin-Sung Kim}
\author[\NVIDIA]{Thien Nguyen}
\author[\NVIDIA]{Pooja Rao}
\author[\keio, \MCC]{Qi Gao}
\author[\keio]{Michihiko Sugawara}
\author[\keio]{Naoki Yamamoto}
\author[\chemUofT,\csUofT,\vecInst,\chemEngUofT,\matSciUofT,\CIFAR,\NVIDIA]{Alan Aspuru-Guzik}
\affil[\chemUofT]{
    Chemical Physics Theory Group, Department of Chemistry, University of Toronto, Toronto, Ontario, Canada}
\affil[\AIST]{%
    Research Center for Emerging Computing Technologies, National Institute of Advanced Industrial Science and Technology (AIST), 1-1-1 Umezono, Tsukuba, Ibaraki, Japan}
\affil[\keio]{%
    Quantum Computing Center, Keio University, 3-14-1 Hiyoshi, Kohoku-ku, Yokohama, Kanagawa, Japan}
\affil[\csUofT]{%
Department of Computer Science, University of Toronto, Toronto, Ontario, Canada}
\affil[\vecInst]{%
    Vector Institute for Artificial Intelligence, Toronto, Ontario, Canada}
\affil[\Harvard]{%
    Department of Physics, Faculty of Arts and Sciences, Harvard University, Cambridge, Massachusetts, USA}
\affil[\SJ]{%
    Department of Structural Biology, St. Jude Children's Research Hospital, Memphis, Tennessee, USA}
\affil[\IMS]{%
    Institute of Medical Science, University of Toronto, Toronto, Ontario, Canada}
\affil[\NVIDIA]{%
NVIDIA, Santa Clara, California, USA
}
\affil[\chemEngUofT]{%
    Department of Chemical Engineering \& Applied Chemistry, University of Toronto, Toronto, Ontario, Canada}
\affil[\matSciUofT]{%
    Department of Materials Science \& Engineering, University of Toronto, Toronto, Ontario, Canada}
\affil[\CIFAR]{%
    Lebovic Fellow, Canadian Institute for Advanced Research, Toronto, Ontario, Canada}
\affil[\MCC]{%
    Mitsubishi Chemical Corporation, Science \& Innovation Center, 1000, Kamoshida-cho, Aoba-ku, Yokohama 227-8502, Japan
}
\begin{document}

\maketitle

\begin{abstract}
\addsec*{Abstract}
We introduce the generative quantum eigensolver (GQE), a new quantum computational framework that operates outside the variational quantum algorithm paradigm by applying classical generative models to quantum simulation. The GQE algorithm optimizes a classical generative model to produce quantum circuits with desired properties. Here, we develop a transformer-based implementation, which we name the generative pre-trained transformer-based (GPT) quantum eigensolver (GPT-QE). We show a proof-of-concept of training and pretraining of GPT-QE applied to electronic structure Hamiltonians, and demonstrate its ability illustrated by surpassing coupled cluster singles and doubles (CCSD) for the strong bond dissociation of the nitrogen molecule and approaching chemical accuracy. We also demonstrate the method on real quantum hardware.
\end{abstract}

\section{Introduction}
The field of quantum computing has experienced a remarkable surge, characterized by rapid advancements in the development of quantum devices. Notably, recent publication reports the experimental realization of quantum computing with 48 logical qubits \cite{bluvstein2023logical}, marking the onset of the early fault-tolerant quantum computing regime. However, despite these advancements, this regime's operational number of gates remains limited. Consequently, it is still unclear how these hardware leaps can be effectively translated into practical advantages in the coming years.

 A decade has passed since some of us introduced the variational quantum eigensolver (VQE) \cite{peruzzo_variational_2014}, which arguably marked a pivotal moment in the field of quantum computing. In VQE, a cost function is minimized by optimizing parameters embedded in a quantum circuit. The variational nature of the algorithm facilitates reducing the depth of the circuit so they can be implemented on near-term devices. Since its introduction, many quantum algorithms employing variational techniques (variational quantum algorithms: VQAs) have been proposed \cite{cerezo_variational_2021,bharti2022noisy}.
However, it has been demonstrated that VQAs encounter several issues, particularly with regards to their trainability for large problem instances \cite{magann_randomized_2023,wang2021noise}. This limitation hinders their competitiveness against classical computers when dealing with problems above a certain size. In this work, we aim to circumvent these shortcomings by constructing a completely different approach that operates outside the VQA paradigm. 
\begin{figure*}[h]
\centering
\includegraphics[width=450pt]{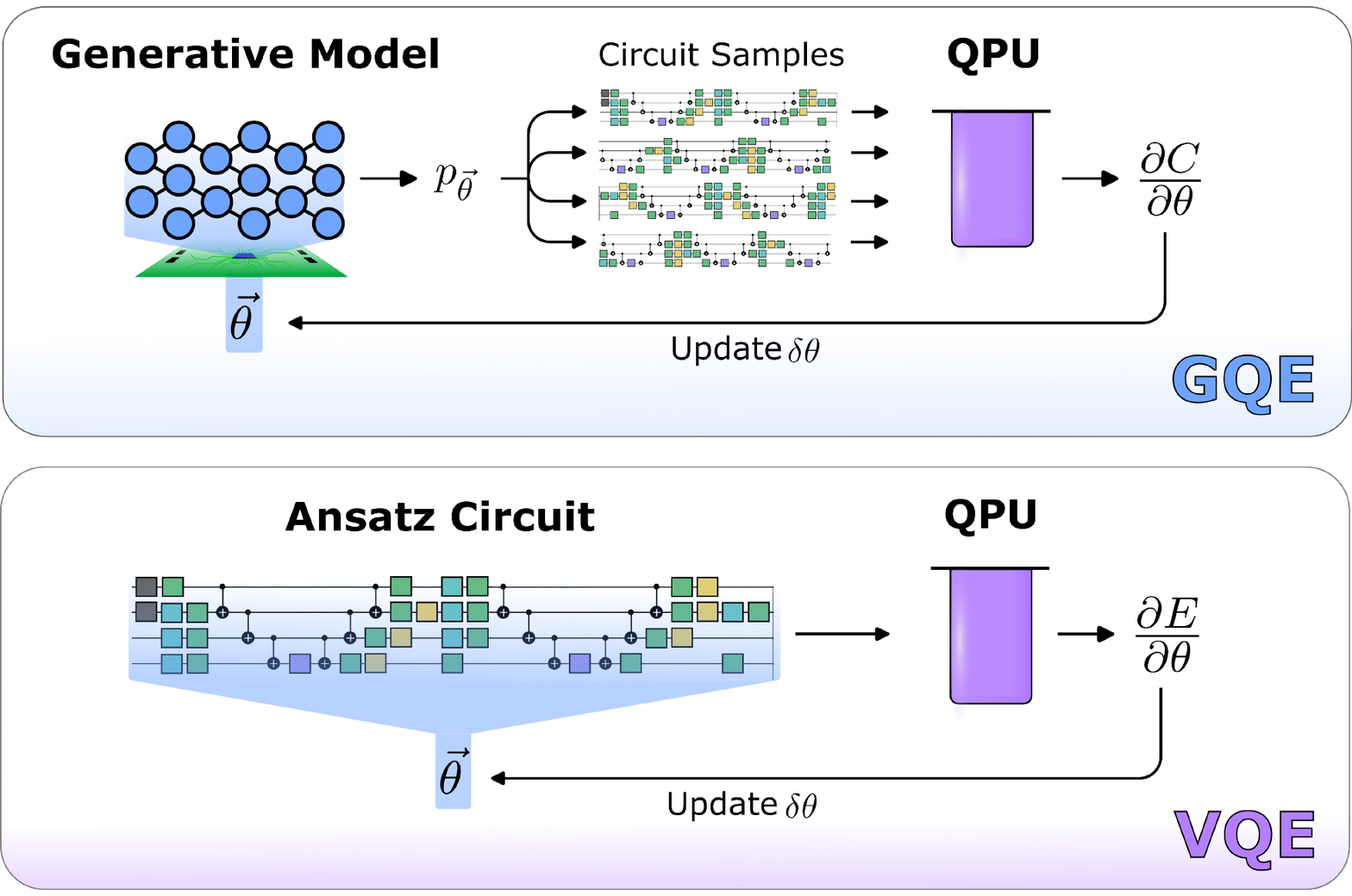}
\caption{Comparison between GQE and VQE.}
\label{fig:gpt_vs_vqe}
\end{figure*}

During the same tumultuous decade, modern machine-learning techniques based on deep neural networks have revolutionized numerous areas. In particular, there has been significant advancement in generative models for natural language processing.
The advent of the Generative Pre-trained Transformer (GPT) \cite{radford_language_2019} marks a milestone in the evolution of artificial intelligence. Forming the basis of Large Language Models (LLMs), GPT-like transformer models have demonstrated exceptional capabilities in understanding and generating human language. Through the simplicity and inherent efficiency of the attention mechanism \cite{vaswani_attention_2017}, transformer models have demonstrated extraordinary performance across a wide array of tasks, showcasing their flexibility and expressivity in a variety of domains (e.g., \cite{radford_language_2019,vaswani_attention_2017,zhao_point_nodate,dosovitskiy_image_2020}). Recent achievements, highlighted by models like Chinchilla \cite{hoffmann_training_2022}, demonstrate how scaling laws in machine learning can inform the efficient allocation of model size for optimized performance, hinting at even greater potential. 

\begin{figure*}[h]
	\centering
	\includegraphics[width=450pt]{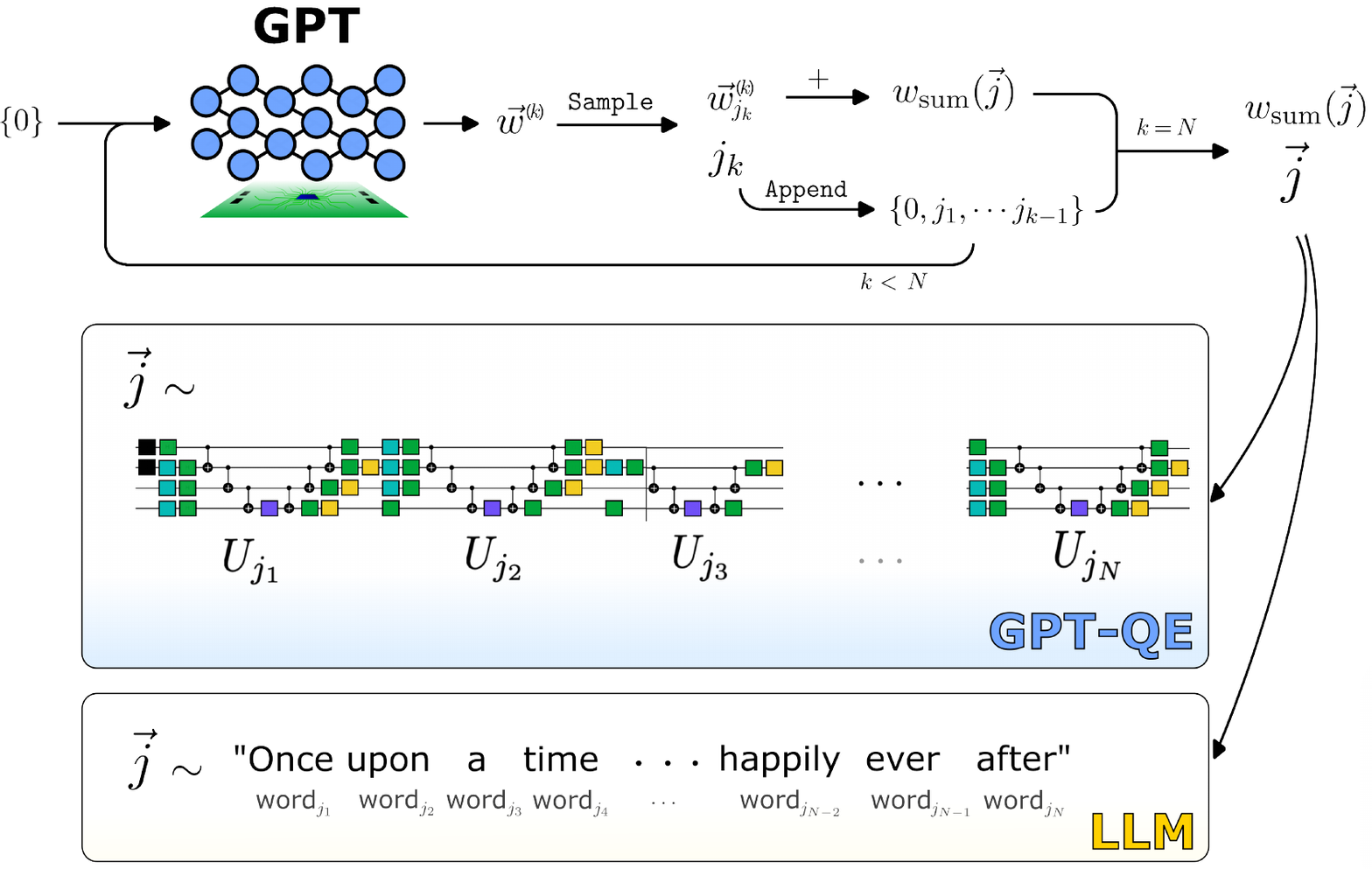}
\caption{Depiction of quantum circuit generation in GPT-QE (GQE, which employs a transformer). We also show the analogy between document generation in Large Language Models (LLMs) and quantum circuit generation in GPT-QE. The details of quantum circuit generation are described in Section~\ref{section:gptqe}.}
\label{fig:concept}
\end{figure*}

Given those significant achievements in classical generative models, incorporating generative models into quantum computing algorithms could be a pivotal step in overcoming the enduring challenges faced in practical quantum computing applications. Therefore, we propose the generative quantum eigensolver (GQE), which takes advantage of classical generative models for quantum circuits. Specifically, we employ a classical generative model —denoted as $p_{\vec{\theta}}(U)$, with $\vec{\theta}$ as parameters and $U$ as a unitary operator—to define a probability distribution for generating quantum circuits. In simpler terms, we sample quantum circuits according to this distribution. We train $p_{\vec{\theta}}(U)$ so that generated quantum circuits are likely to have desirable properties. We emphasize that, unlike VQA and its variants, no parameters are embedded in the quantum circuit in GQE; notably, and importantly for scalability, \textsl{ all} the optimizable parameters are in the classical generative model (Fig.~\ref{fig:gpt_vs_vqe}). We note that some previous works utilize a generative model for generating parameters in VQE \cite{verdon2019learning,kim2023preparation,yang2023maximising}, but in GQE, the whole circuit structure is determined by a generative model.

In designing the generative model for quantum circuits generation, we focus on the transformer architecture \cite{vaswani_attention_2017}, which has achieved significant success as the backbone of large language models. 
We can describe the implementation of GQE with a transformer by using an analogy between natural language documents and quantum circuits (Fig.~\ref{fig:concept}). 
 For a given operator pool, defined by a set of unitary operations $\{U_j\}_{j=1}^L$ (vocabulary), the transformer generates the sequence of indices $j_1 \dots j_N$ corresponding to the unitary operations $U_{j_1} \dots U_{j_N}$ (words) and constructs the quantum circuit $U_N(\vec{j})=U_{j_N}\dots U_{j_1}$ (document). The rules for generating indices (grammar) are trained so that a cost value calculated by quantum devices decreases. 
 GQE with a transformer is also able to be pre-trained. If we have a dataset given as pairs of index sequences and cost values: $\{\vec{j}_m, C(\vec{j}_m)\}_{m=1}^M$ (document dataset), we can pre-train the transformer without running quantum devices, as shown in Section~\ref{section:methods}. 
Hence, we give the GQE with transformer the name of generative pre-trained transformer-based quantum eigensolver (GPT-QE). 

To demonstrate the effectiveness of our fundamentally new GPT-QE approach, this paper demonstrates solving the molecular electronic ground state search problem, which has significant utility in applications including drug discovery \cite{heifetz_quantum_2020,lam_applications_2020} and materials science \cite{agarwal_discovery_2021}. Our results demonstrate that GPT-QE outperforms coupled cluster singles and doubles (CCSD) methods in the strong bond dissociation regime of the nitrogen molecule. 

An important distinction from traditional VQE approaches is the potential for incorporating conditional inputs into the generative model. While this paper does not directly demonstrate this capability, the transformer architecture enables the integration of domain knowledge or problem-specific constraints as conditional inputs during circuit generation. This represents a fundamentally different paradigm compared to VQE, where the circuit structure is typically fixed and only parameters are optimized. We discuss this capability and its implications in Section~\ref{section:gptqe}. Following the initial preprint of this work, subsequent research has successfully demonstrated this conditional input capability for combinatorial optimization problems \cite{minami2025generative}, validating the potential of this approach.

We note that previous literature \cite{liang2023unleashing,krenn2023artificial,jaouni2023deep} proposes methods for constructing the structure of quantum circuits using machine learning, especially reinforcement learning. Reinforcement learning approaches tend to require access to a large number of intermediate quantum states in the circuit to determine each action (the next quantum gate to be generated), which leads to an increase in the number of required measurements as the number of gates increases. 
Conversely, GPT-QE does not require any intermediate measurements, thus potentially significantly reducing the measurement cost when running the algorithm.

The rest of the paper is organized as follows. In Section~\ref{section:methods}, we describe the details of GQE. Particularly, we construct GPT-QE and describe its training and pre-training scheme. Section~\ref{section:result} is dedicated to demonstrating the training and pre-training schemes by using the ground state search for the electronic structure Hamiltonians as a benchmark. It also presents the results of GPT-QE on a noisy real device, with particular focus on the effectiveness of error mitigation techniques in improving the training process. In Section~\ref{section:conclusion}, we summarize what we know about the algorithm so far and suggest future directions of exploration.

\section{Methods}
\label{section:methods}

\subsection{Generative Quantum Eigensolver}
The generative quantum eigensolver (GQE) is an algorithm to search for the ground state of a given Hamiltonian $\hat{H}$. Particularly, we focus on the electronic structure problem, where the Hamiltonian is written as the weighted sum of the tensor products of Pauli operators $\hat{P}_{\ell}$: $\hat{H} = \sum_{\ell} h_{\ell} \hat{P}_{\ell}$.
To construct the approach of GQE, we first illustrate our formulation of the generative model of quantum circuits. 

We prepare the operator pool $\mathcal{G}=\{U_j\}_{j=1}^L$, where $U_j$ is a unitary operator and $L$ is the size of the operator pool.
One of the possible choices for the operator pool is a set of time evolution operators: $\{e^{i \hat{P}_j t_j}\}_{j=1}^L$, which is what we use in our numerical experiments. The detailed configuration of our operator pool is described in Appendix~\ref{app:experiment_details}.
Given a sequence length $N$, we sample the sequence $\vec{j} = \{j_1, \ldots, j_N\}$ according to the parameterized probability distribution $p_N(\vec{\theta}, \vec{j})$, where $\vec{\theta} = \{\theta_p\}_{p=1}^P$ are optimizable parameters. Using the sequence $\vec{j}$, we construct the quantum circuit $U_N(\vec{j}) = U_{j_N} \cdots U_{j_1}$. We call $p_N(\vec{\theta}, \vec{j})$ the generative model of quantum circuits. In the rest of the paper, we omit the variable $\vec{\theta}$ for simplicity. The process of sampling the sequence $\vec{j}$ according to $p_N(\vec{j})$ and constructing the quantum circuit $U(\vec{j})$ is simply referred to as ``sampling the quantum circuit $U(\vec{j})$ according to $p_N(\vec{j})$''.

We construct GQE to search for the ground state of $\hat{H}$ using the generative model of quantum circuits $p_N(\vec{j})$. 
The objective of the problem we target is finding $\vec{j}^{\ast} := \argmin_{\vec{j}}E(\hat{H}, \vec{j})$ with 
\begin{equation}
	E(\hat{H}, \vec{j}) := \tr\left( \hat{H} U_N(\vec{j}) \rhoinit U_N(\vec{j})^{\dagger} \right), 
\end{equation}
where $\rhoinit$ is a fixed initial quantum state. 
In the following, we omit the variable $\hat{H}$ depending on the context for simplicity. 

We note that we need to select the operator pool $\mathcal{G}$ to be expressive enough, 
so that $E( \vec{j}^{\ast})$ is close enough to the ground state energy. 
It should also be noted that we can select $\mathcal{G}$ to accommodate the native operations and topology of each quantum device, or to reflect domain-specific insights.
We illustrate a specific choice for a chemistry-inspired operator pool in our numerical experiment. We train $p_N(\vec{j})$ so that the generated quantum circuits $U_N(\vec{j})$ are likely to produce a low energy quantum state. We call the approach to optimize $p_N(\vec{j})$ as the generative quantum eigensolver. 

We now emphasize the difference between VQE and GQE. As shown in Fig.~\ref{fig:gpt_vs_vqe}, in VQE, we embed parameters in the quantum circuit and optimize them to minimize the energy associated with the generated quantum state. In contrast, all parameters in GQE are embedded in the generative model $p_N(\vec{j})$. 
Consequently, the cost function landscapes in GQE and VQE are different; considering the success of training large models with DNN \cite{garipov2018loss,allen2019convergence,zhou2019sgd}, we expect
that GQE can potentially address certain issues of trainability in VQE by exploiting the different landscape, which has been now moved onto the classical computer.

An advantage of the GQE approach is that we are free to choose the generative model $p_N(\vec{j})$ from a very rich potential set of families of generative models stemming from the field of machine learning. 
Another advantage is the ability to incorporate contextual inputs, enabling the generation of problem-specific quantum circuits based on classical information such as molecular geometry or optimization constraints, as demonstrated in combinatorial optimization applications \cite{minami2025generative}.

\subsection{GPT Quantum Eigensolver}
\label{section:gptqe}
We construct a specific GQE algorithm using the transformer architecture and provide its training scheme. 
As we will show later, the approach also involves pre-training; therefore, we call the method Generative pre-trained transformer-based quantum eigensolver (GPT-QE). 
In the following, we describe how the transformer generates quantum circuits. Then, we construct the training/pre-training scheme of GPT-QE.

\subsubsection{Quantum circuits generation in GPT-QE}
\label{section:gptqe-sampling}
The original transformer, introduced in \cite{vaswani_attention_2017}, targets neural machine translation,  where the model consists of an encoder for the input language and a decoder for the targeted language. In quantum circuit generation, we focus on the decoder-only transformer inspired by GPT-2 \cite{radford_language_2019}, developed for more general generative tasks. In the following, we refer to a decoder-only transformer simply as the transformer.

The sequence generation using the transformer can be written as repetitions of (i) calculating the logit (logarithmic probability) with which each token is generated and (ii) sampling a token according to the corresponding probability distribution. We write the function for the logit calculation as \texttt{GPT} and the function for sampling as  \texttt{Sample}. 

The function \texttt{GPT} takes the variable-length inputs $\vec{j}^{(k)} = \{j_1, \ldots, j_k \}$, where each element takes an integer value between $1$ and $L$ (the size of the operator pool $\mathcal{G}$). 
Then it outputs a sequence of logits 
$W^{(k)} = \{\vec{w}^{(1)}, \ldots, \vec{w}^{(k)} \}$ with the same length, where each logit $\vec{w}^{(r)}$ is a real vector of size $L$. We note that \texttt{GPT} has optimizable parameters $\vec{\theta}$ included in the transformer's architecture, which is not explicitly written in the notation \texttt{GPT} for simplicity. 

The function \texttt{GPT} follows the methodology outlined in \cite{radford_language_2019}. Here, we present a concise overview of \texttt{GPT}'s operation:
Initially, \texttt{GPT} converts each token in the input sequence into a unique embedding, represented as a vector. These embeddings are transformed by means of multiple attention layers. Each attention layer takes a sequence of vectors as input, with the first layer's input being the embeddings themselves. The output of each attention layer is also a sequence of vectors, maintaining the same sequence length and vector dimension as the input. This consistency allows the output of one layer to serve as the input for the subsequent layer. The final attention layer's output is converted into a sequence of logits, constituting the output of \texttt{GPT}.
Given an attention layer's input $\{\vec{v}_1, \cdots, \vec{v}_k\}$, the output can be expressed as $\{\vec{a}_1, \cdots, \vec{a}_k\}$, where each $\vec{a}_r$ represents the attention with the same dimension as $\vec{v}_r$. The attention  $\vec{a}_r$ encapsulates the relationship between $\vec{v}_r$ and other elements of the input. Notably, through the process of causal masking, $\vec{a}_r$ depends solely on preceding elements, i.e., $\{\vec{v}_{r^{\prime}} \}_{r^{\prime}<r}$. The standard implementation of attention computation, as employed here, involves running multiple attention mechanisms in parallel, and their outputs are combined into $\{\vec{a}_1, \cdots, \vec{a}_k\}$ through a weighted average (multi-head attention).
For a more comprehensive explanation, see \cite{radford_language_2019}.

\begin{figure*}[h]
\centering
\includegraphics[width=500pt]{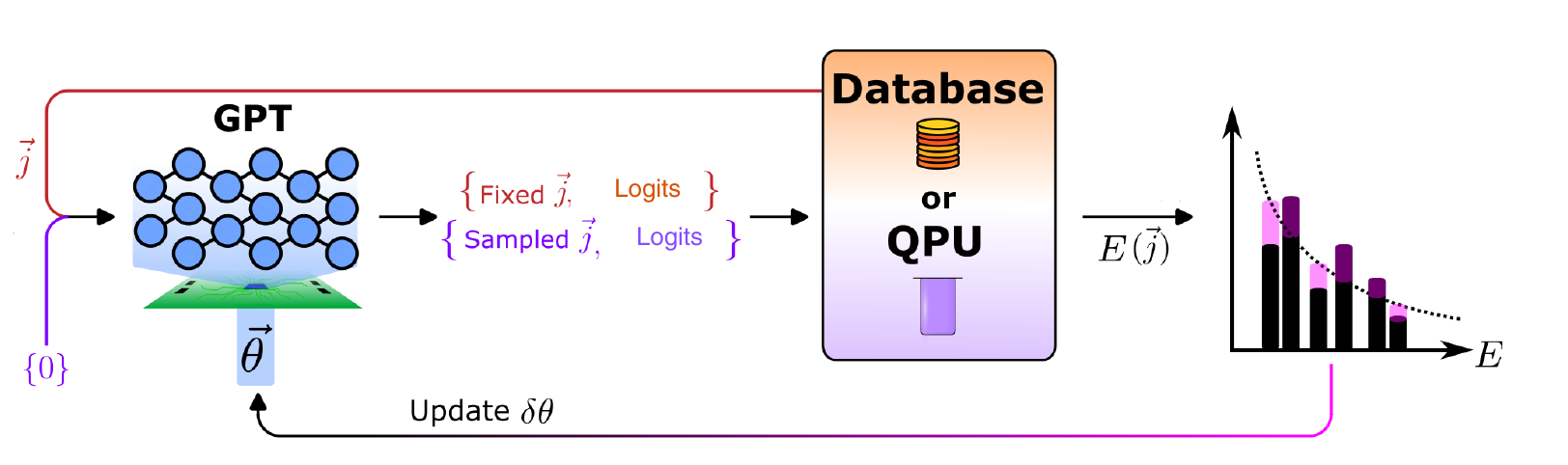}
\caption{Overview of the training and pre-training scheme in GPT-QE. }
\label{fig:methodGPTQE}
\end{figure*}

The function \texttt{Sample} is a stochastic function that takes a logit $\vec{w} = \{w_j\}_{j=1}^L$ as its input and returns one of the tokens $j \in \{1,\ldots,L\}$. The probability that the token $j$ is sampled is proportional to $e^{-\beta w_j}$, 
with $\beta > 0$ a hyper-parameter we can choose. 
This can be understood as each $j$ being sampled according to the energy $w_j$ and the inverse temperature $\beta$ as in statistical mechanics.
For simplicity, we omit the variable $\beta$ from the function \texttt{Sample}. 

We define the generative model of the quantum circuits in GPT-QE by the procedure to obtain a sequence $\vec{j}$ using \texttt{GPT} and \texttt{Sample}:
\begin{itemize}
	\item In the first step, we sample the token $j_1$ with the fixed input $\{0\}$:
	\begin{align}
	W^{(1)}  &=   \texttt{GPT}(\{0\}), \nonumber \\
	\vec{w}^{(1)} &= W^{(1)}_1, \nonumber \\
	j_1 &= \texttt{Sample}(	\vec{w}^{(1)}), \nonumber 
	\end{align}
	where $W^{(1)}_1$ denotes the first element of $W^{(1)}$.
	\item In the second step, we sample the token $j_2$ with the input $\{0, j_1\}$:
		\begin{align}
	W^{(2)}  &= \texttt{GPT}(\{0, j_1\}), \nonumber \\
	\vec{w}^{(2)} &= W^{(2)}_2, \nonumber \\
	j_2 &= \texttt{Sample}(\vec{w}^{(2)})\nonumber .
	\end{align}
	\item In the $k$-th step, we sample the token $j_k$ with the input $\{0, j_1, \ldots, j_{k-1}\}$:
		\begin{align}
	W^{(k)} &= \texttt{GPT}(\{0, j_1, \ldots, j_{k-1}\}), \nonumber \\
	\vec{w}^{(k)} &= W^{(k)}_k, \nonumber \\
	j_k &= \texttt{Sample}(\vec{w}^{(k)}).\nonumber 
	\end{align}
\end{itemize}
After $N$ steps, we obtain a sequence of tokens $\vec{j} = \{j_1, \ldots, j_N\}$ with length $N$. 

We can readily show that the probability that $\vec{j}$ is sampled is proportional to 
$\exp\left(-\beta w_{\textrm{sum}}(\vec{j})\right)$, where 
\begin{equation}
	w_{\textrm{sum}}(\vec{j}) := \sum_{k=1}^{N} w_{j_k}^{(k)}.
\end{equation}
Therefore, the generative model in GPT-QE is 
\begin{equation}
	p_N(\beta, \vec{j}) = \frac{\exp\left(-\beta w_{\textrm{sum}}(\vec{j})\right)}{\mathcal{Z}}, 
\end{equation}
where $\mathcal{Z}$ is a normaliziation factor and we write the hyper-parameter $\beta$ explicitly.

\subsubsection{Training}

The training process in GPT-QE involves two main phases: data collection through circuit sampling and model updates using batched data from the replay buffer. The replay buffer is a technique commonly used in reinforcement learning \cite{lin1992self} to store and reuse past experiences for more efficient learning. We implement training with a replay buffer of fixed size $N_{\text{buffer}}^{\text{max}}$ that operates as a first-in-first-out (FIFO) queue to store and manage previously sampled circuits and their corresponding energies.

In each training epoch, we collect $N_{\text{sample}}$ new circuit samples through the process described in section \ref{section:gptqe-sampling}. We evaluate the corresponding energies of circuits using the quantum device (or its simulator). These sequence-energy pairs are then stored in the replay buffer using a FIFO mechanism. The replay buffer contains a dataset $\mathcal{Q} = \{\vec{j}_{m}, E(\vec{j}_{m})\}_{m=1}^{N_{\text{buffer}}}$ where $N_{\text{buffer}}$ represents the current number of entries in the buffer. When the buffer reaches its maximum capacity $N_{\text{buffer}}^{\text{max}}$, the oldest entries are automatically removed to make room for new data. Note that when $N_{\text{buffer}}^{\textrm{max}} = N_{\text{sample}}$, this means the newly collected data is used immediately and then disposed of.

The model update phase utilizes data from the replay buffer to improve the model's performance. During this phase, the $N_{\text{buffer}}$ data points in the replay buffer are processed in batches of size $N_{\text{batch}}$, and this process is repeated for $N_{\text{iter}}$ iterations. In each iteration, the loss function is computed using the current batch of data points, and the model parameters are updated through gradient descent to minimize the loss.  The loss function is designed so that logits corresponding to high-energy circuits become smaller, while logits corresponding to low-energy circuits become larger. This encourages the model to generate circuits with lower energies. As our specific construction of the loss functions, we use the logit matching loss and Group Relative Policy Optimization (GRPO) \cite{GRPO2024} loss in our experiment, as introduced in  Appendix~\ref{app:loss_functions}. 

The nature of the training in the GPT-QE approach naturally enables pre-training capabilities. By storing previous learning experiences in the replay buffer, we can leverage this accumulated data to pre-train models for new tasks or configurations, providing a foundation for learning across different quantum optimization problems. The pre-training process is entirely classical, eliminating the need to use quantum devices as long as datasets are available.

In order to reuse the data in the replay buffer for a different problem, the energies of the replay buffer should be translated to the energies of the new Hamiltonian $H$. For this purpose, we utilize that, when computing on quantum devices, we evaluate the expectation values for each Pauli basis, from which the expectation value of the target Hamiltonian is constructed. These individual Pauli expectation values can be stored and reused for estimating different observables, significantly improving computational efficiency. In particular, it allows the energy of $H^\prime$ to be estimated if it contains the same Pauli strings as the original Hamiltonian. The most prominent example of this approach is the electronic structure Hamiltonian, where the same set of Pauli expectation values can be used to estimate energies for different molecular geometries. This technique is detailed in Appendix~\ref{app:coefficient_reweighting} and investigated in Section~\ref{section:pretraining} (see also \cite{self_variational_2021}). 

Above, we describe the case where GPT-QE itself constructs the dataset. However, it is also possible that the dataset could be generated from other algorithms, e.g., tensor network calculations and VQE, if we can (approximately) convert the data obtained in those algorithms to a data format acceptable for GPT-QE. Following our initial submission, subsequent work has demonstrated this approach in quantum combinatorial optimization \cite{tyagin2025qaoa}.

\subsubsection{Conditional generalization}
\label{section:transfer}
A key advantage of using generative models is the ability to incorporate conditional inputs, enabling the generation of problem-specific quantum circuits based on classical information such as molecular geometry or optimization constraints. We can define a conditional GPT model as $\texttt{GPT}(\vec{\Delta})$, where $\vec{\Delta}$ represents the conditional input parameters (e.g., molecular geometry, bond lengths, or other problem-specific constraints). This model can generate quantum circuits that are tailored to specific parameter values $\vec{\Delta}$, and the corresponding cost can be evaluated using the Hamiltonian $\hat{H}(\vec{\Delta})$ for that particular configuration. When a new parameter configuration $\vec{\Delta}_{\text{new}}$ is encountered, full retraining is likely not required—the model can immediately generate appropriate quantum circuits for the new configuration using the learned conditional mapping, though fine-tuning remains an option for further optimization. This capability opens up new possibilities for generalization across electronic structure problems, where we can leverage knowledge gained from one system to improve performance on related systems. While this paper does not provide actual demonstrations of these generalization capabilities, we outline the promising directions and potential applications.

One approach is config-to-config transfer, where we train the model on a set of molecular geometries and then apply it to new geometries of the same molecule. This would extend the pre-training framework by utlilizing a single model across multiple geometries, rather than re-using data for training a separate model for each geometry.
Another promising direction is molecule-to-molecule transfer, where we leverage datasets generated from training on one molecule to improve performance on chemically related molecules. As mentioned, both approaches require extending the model to incorporate molecular information as conditional inputs. While this represents a more challenging extension, it could enable the development of general-purpose quantum circuit generators that can adapt to different molecular systems. 

The primary challenge for the molecule-to-molecule transfer lies in uniformly treating molecules with varying numbers and types of molecular orbitals. For instance, there is not a straightforward mapping between the molecular orbitals of $\texttt{H}_2$ and those of $\texttt{H}_2\texttt{O}$ as determined by Hartree-Fock calculations. Therefore, the gate sequence in the calculation of $\texttt{H}_2$ does not have an a priori correspondence with that in $\texttt{H}_2\texttt{O}$. Finding transferable representations of electronic structure information that are useful for machine learning applications is a vibrant research field on its own \cite{qiao_informing_2022, khan_kernel_2023}. One potential solution to this challenge in our setting is to write the molecular Hamiltonian using the basis of atomic orbitals instead of that of molecular orbitals. More specifically, we define the creation and the annihilation operators for each atomic orbital and correspond them with qubits. Then, a gate sequence for $\texttt{H}_2$ has an analogous physical meaning when applied to the portion corresponding to hydrogen atoms in $\texttt{H}_2 \texttt{O}$. This approach may facilitate effective molecule-to-molecule transfer, particularly between molecules with the same atomic composition. Alternatively, one could devise a quantum circuit that couples molecular fragments encoded initially in independent qubits. In this manner, the optimized gates for the fragments can be invoked as a starting point for the full molecular circuit.

Note that following our initial preprint, subsequent work has demonstrated the effectiveness of conditional generalization approaches in quantum combinatorial optimization, where conditional inputs have been successfully used to generate problem-specific quantum circuits \cite{minami2025generative,tyagin2025qaoa}.
These developments validate the potential of our generative approach for conditional generalization across different quantum optimization problems.

\begin{figure*}[ht]
    \centering
    \captionsetup[subfigure]{labelformat=empty}
    \begin{tabular}{cc}
        \begin{subfigure}[b]{0.45\textwidth}
            \centering
            \includegraphics[width=\textwidth]{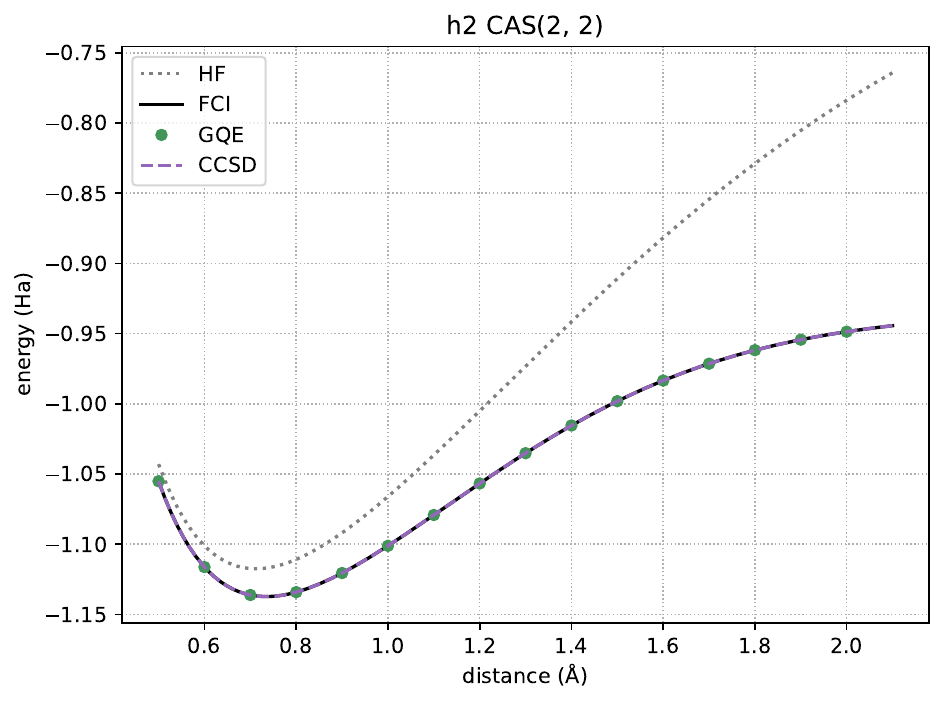}
            \caption{$\texttt{H}_2$ (4 qubits)}
        \end{subfigure} &
        \begin{subfigure}[b]{0.45\textwidth}
            \centering
            \includegraphics[width=\textwidth]{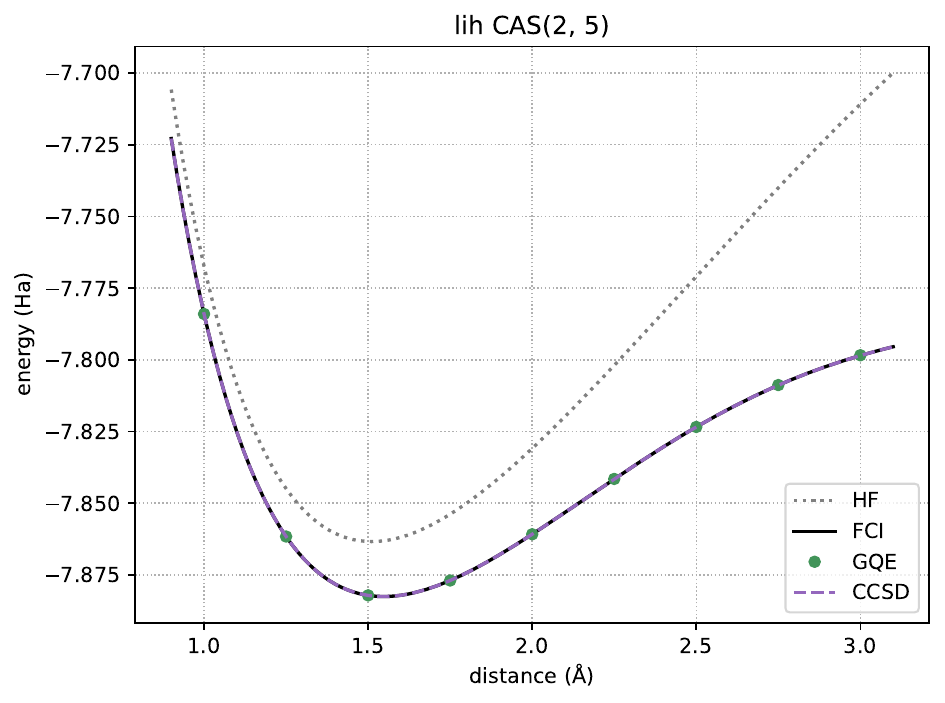}
            \caption{$\texttt{LiH}$ (10 qubits)}
        \end{subfigure} \\
        \begin{subfigure}[b]{0.45\textwidth}
            \centering
            \includegraphics[width=\textwidth]{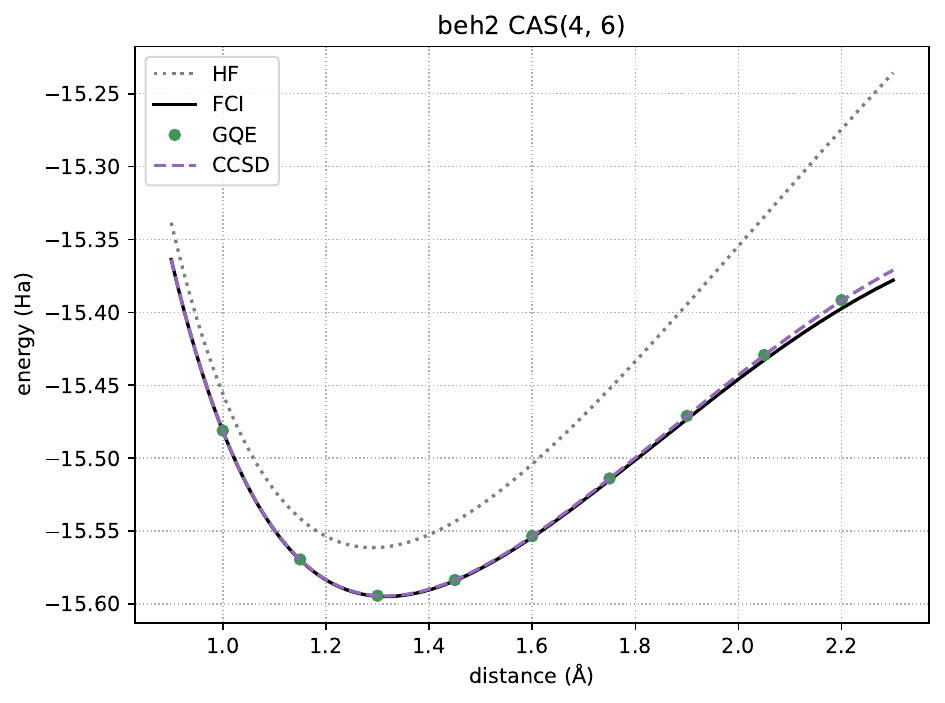}
            \caption{$\texttt{BeH}_2$ (12 qubits)}
        \end{subfigure} &
        \begin{subfigure}[b]{0.45\textwidth}
            \centering
            \includegraphics[width=\textwidth]{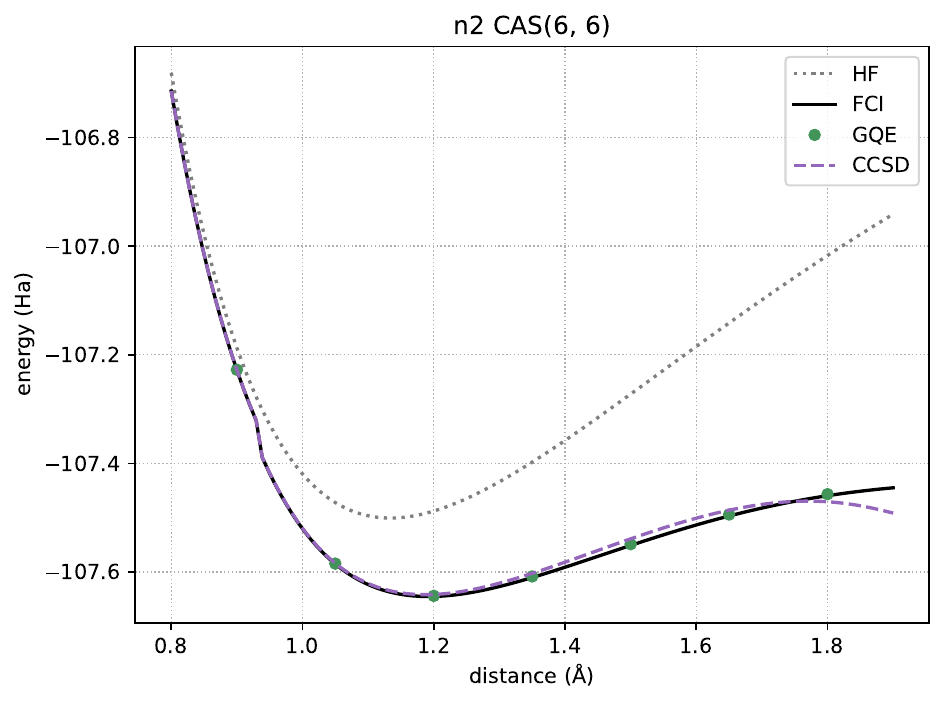}
            \caption{$\texttt{N}_2$ (12 qubits)}
        \end{subfigure}
    \end{tabular}
    \caption{Training results for each molecule showing energy vs. bond length plots with GQE results (green points), HF energy (gray dotted line), CCSD energy (blue dashed line), and FCI energy (black line). The number of tokens (circuit length) is set to 10 for $\texttt{H}_2$, 40 for $\texttt{LiH}$, 60 for $\texttt{BeH}_2$, and 100 for $\texttt{N}_2$. Note that there is a jump in the FCI calculation for $\texttt{N}_2$ around 0.93-0.94 Å, which is due to a change in the active space selection in the CAS calculation. This does not affect the objectives or conclusions of this experiment since it affect all approaches equally.}
    \label{figure:training}
\end{figure*}

\begin{figure*}[ht]
    \centering
    \captionsetup[subfigure]{labelformat=empty}
    \begin{tabular}{cc}
        \begin{subfigure}[b]{0.45\textwidth}
            \centering
            \includegraphics[width=\textwidth]{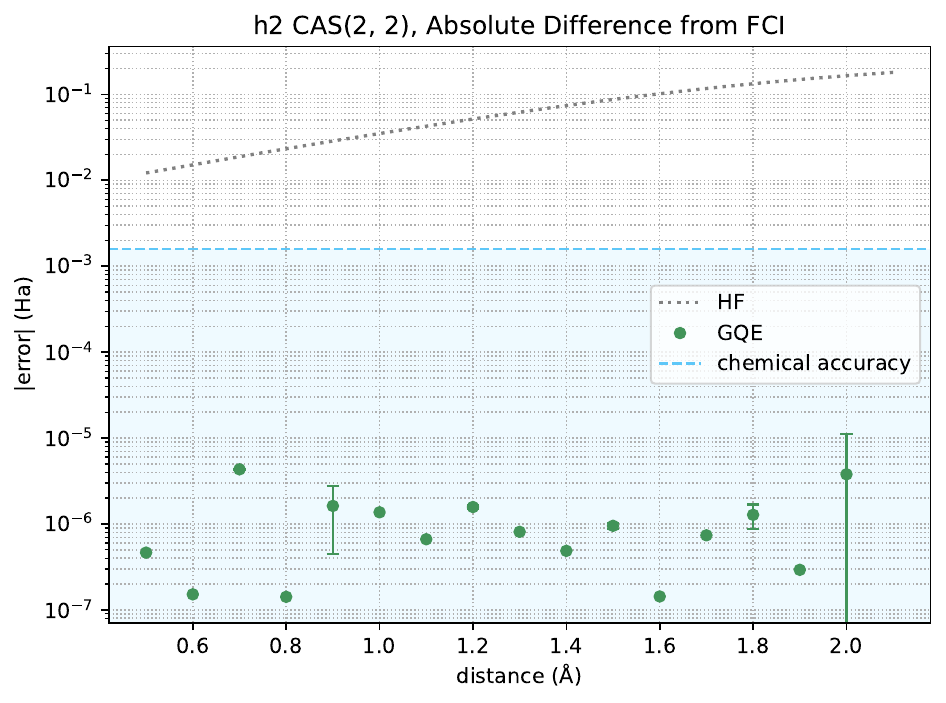}
            \caption{$\texttt{H}_2$ (4 qubits)}
        \end{subfigure} &
        \begin{subfigure}[b]{0.45\textwidth}
            \centering
            \includegraphics[width=\textwidth]{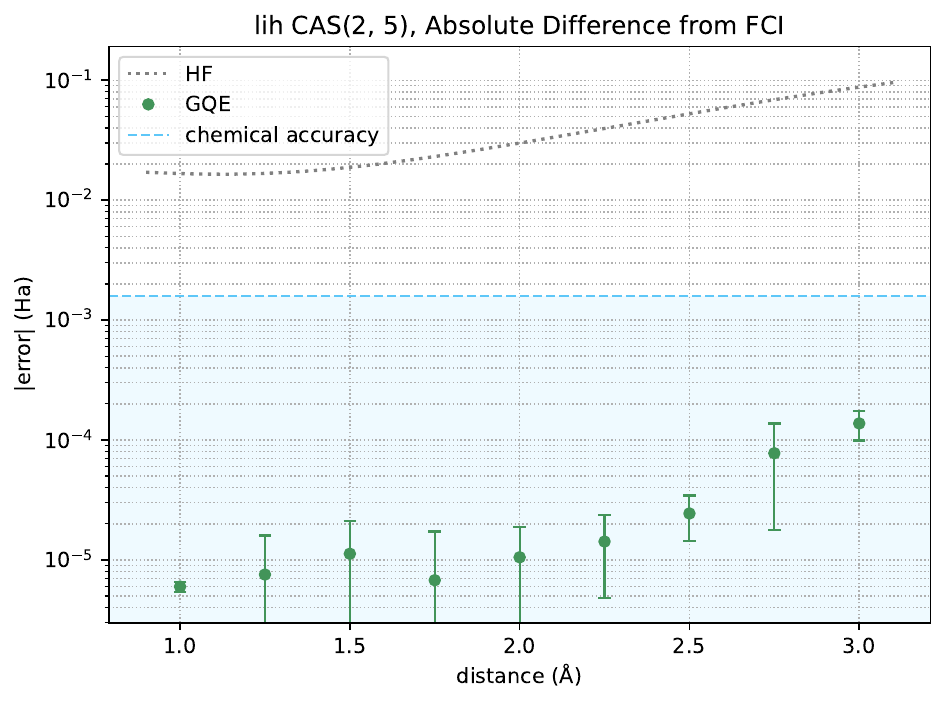}
            \caption{$\texttt{LiH}$ (10 qubits)}
        \end{subfigure} \\
        \begin{subfigure}[b]{0.45\textwidth}
            \centering
            \includegraphics[width=\textwidth]{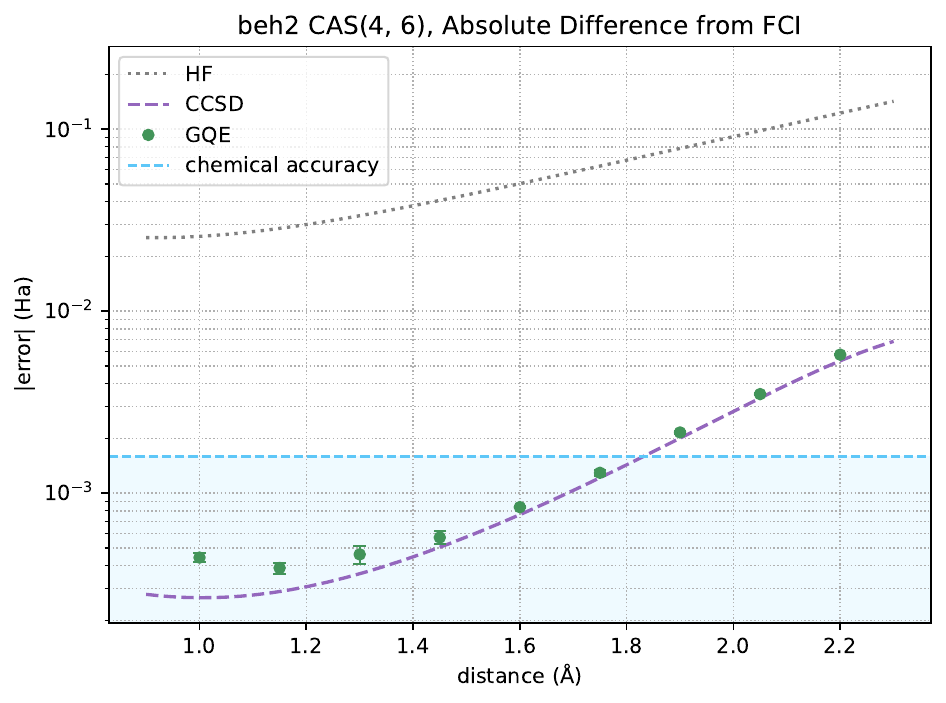}
            \caption{$\texttt{BeH}_2$ (12 qubits)}
        \end{subfigure} &
        \begin{subfigure}[b]{0.45\textwidth}
            \centering
            \includegraphics[width=\textwidth]{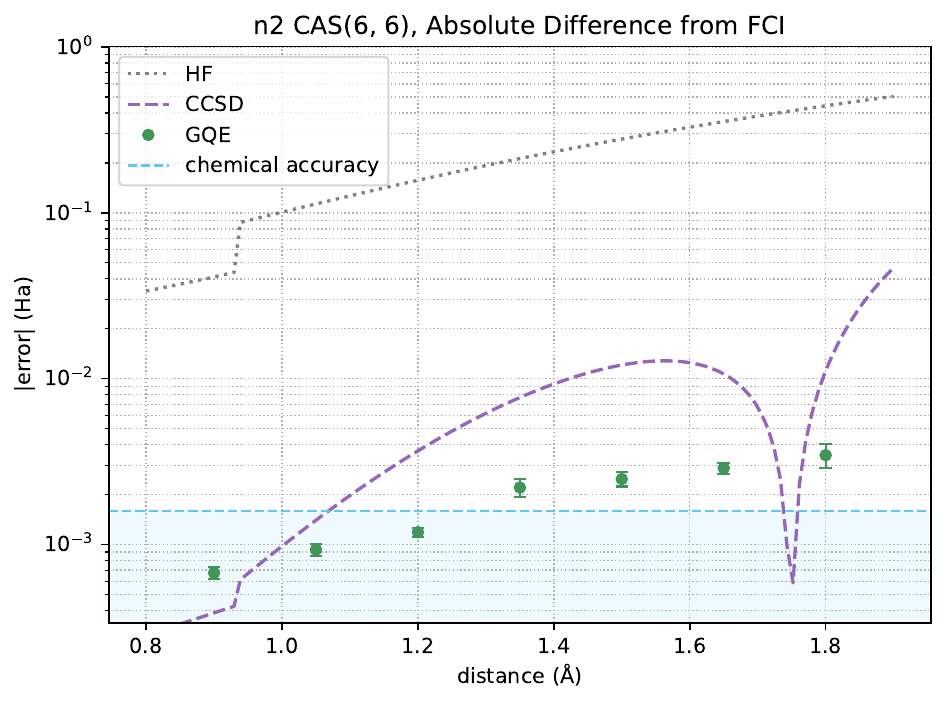}
            \caption{$\texttt{N}_2$ (12 qubits)}
        \end{subfigure}
    \end{tabular}
    \caption{Absolute error from FCI on a logarithmic scale for each molecule, with chemical accuracy threshold (blue dashed line) and HF/CCSD reference lines. The circuit complexity increases with molecular size, requiring longer sequences for more complex molecules. Note that for H$_2$ and LiH, CCSD is extremely close to FCI, so it is not plotted here. Additionally, the kink in N$_2$ around 1.75 Å is due to a sign change in the error.}
    \label{figure:training-detail}
\end{figure*}

\begin{figure}[h]
\centering
\includegraphics[width=0.45\textwidth]{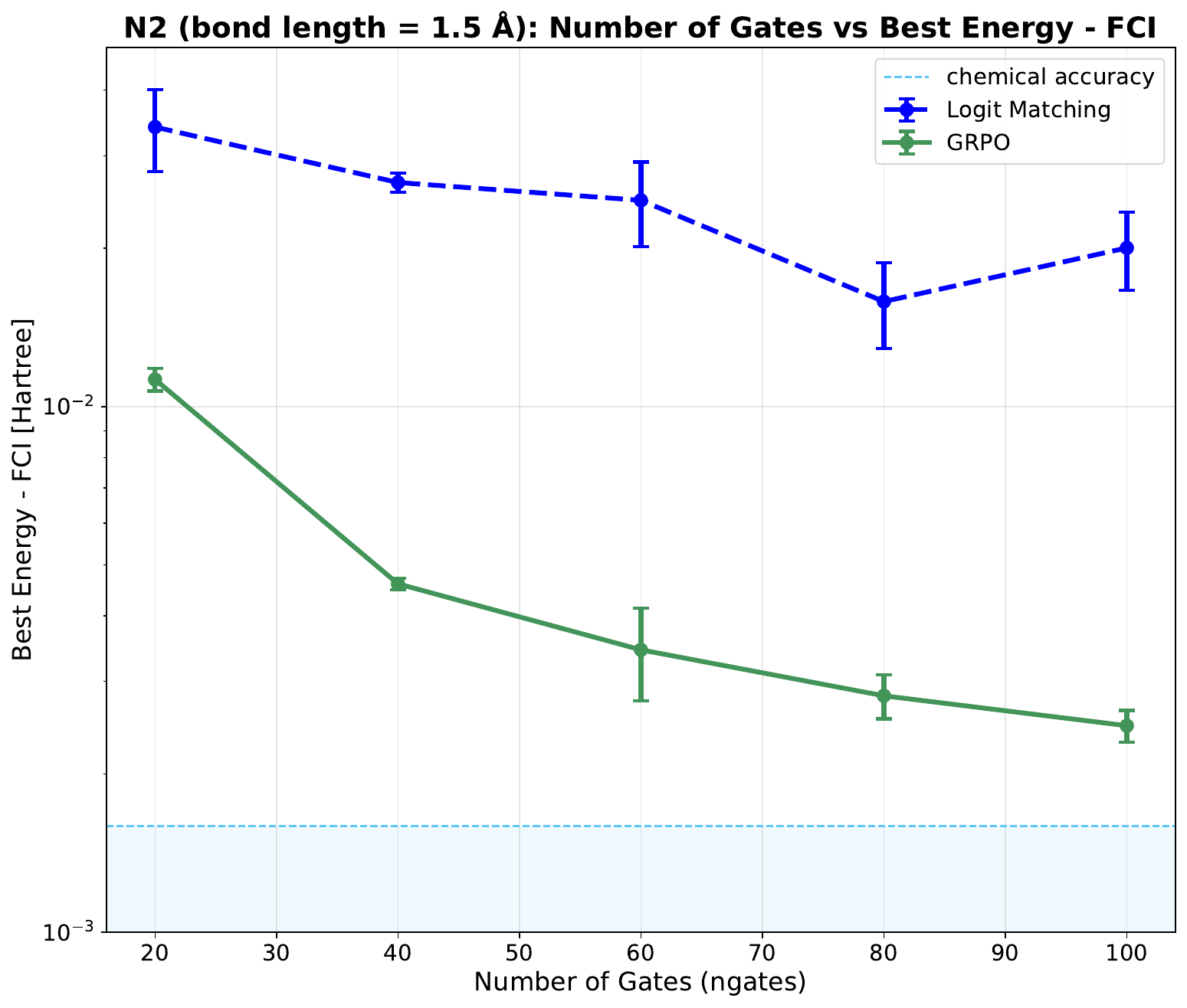}
\caption{Comparison of GRPO and Logit Matching loss functions for N$_2$ at bond length 1.5 Å. The plot shows the best energy relative to the ground truth energy as a function of the number of gates (circuit depth). }
\label{fig:gate_comparison}
\end{figure}

\section{Results}
\label{section:result}

In this section, we investigate the effectiveness of training in GPT-QE for approximating ground states using electronic structure Hamiltonians. We use the molecular Hamiltonians of $\texttt{H}_2$, $\texttt{LiH}$, $\texttt{BeH}_2$, and $\texttt{N}_2$ in the \texttt{sto-3g} basis for this purpose. The more detailed process to generate molecular systems is described in Appendix~\ref{app:molecule_generation}.

The configuration of the GPT-QE model is as follows. Our operator pool consists of Pauli time evolution operators derived from chemically inspired operators based on unitary coupled-cluster single and double excitations (UCCSD), with detailed configuration provided in Appendix~\ref{app:operator_pool}. Regarding the transformer model, we employ a configuration based on GPT-2 \cite{radford_language_2019}, modified to feature 6 attention layers and 6 attention heads. The hyper-parameter $\beta$ is updated in each epoch of training using the strategy described in Appendix~\ref{app:beta_adaptive}. The initial state, $\rhoinit$, is set to the Hartree-Fock state.

In this study, we utilized CUDA Quantum \cite{cuda-quantum} to execute quantum chemistry experiments. CUDA Quantum is distinguished as an open-source programming model and platform, integrating quantum processing units (QPUs), CPUs, and GPUs seamlessly. This integration makes it an ideal choice for workflows that require diverse computing capabilities, as demonstrated in the GPT-QE application we considered. CUDA Quantum facilitates kernel-based programming and is compatible with both C++ and Python, which we employed in our research. The algorithm was executed using NVIDIA A100 GPUs on NERSC's Perlmutter, an HPE Cray Shasta-based heterogeneous system with 1,792 GPU-accelerated nodes.

\subsection{Training Performance}
\label{section:training}

For the training experiments, we set the replay buffer size to $N_{\text{buffer}}^{\textrm{max}} = 1,000$, the batch size to $N_{\text{batch}} = 50$, the number of circuits generated per epoch to $N_{\text{sample}} = 50$, and the number of iterations per epoch to $N_{\text{iter}} = 5$. Before starting the training process, we first collected 200 samples to populate the replay buffer, ensuring a sufficient initial dataset for stable training. The training was conducted for different numbers of epochs depending on the molecular complexity: $\texttt{H}_2$ was trained for 200 epochs, $\texttt{LiH}$ for 1,000 epochs, and $\texttt{BeH}_2$ and $\texttt{N}_2$ for 1,500 epochs. For the loss function, we use Group Relative Policy Optimization (GRPO) \cite{GRPO2024} loss described in Appendix~\ref{app:loss_functions}. All results presented in this section represent the average (standard deviation) over 3 independent trials.

Figure~\ref{figure:training} presents the minimum energies found during the training process for four molecular systems ($\texttt{H}_2$, $\texttt{LiH}$, $\texttt{BeH}_2$, and $\texttt{N}_2$), showing the energy as a function of bond length. The GQE results are displayed as green points, representing the lowest energy values discovered throughout the training epochs, while reference methods are shown as lines: Hartree-Fock (HF) energy in gray dotted lines, coupled cluster singles and doubles (CCSD) energy in blue dashed lines, and full configuration interaction (FCI) energy in black solid lines. The results demonstrate that GQE successfully captures the qualitative behavior of the potential energy surfaces across different molecular systems, with the generated quantum circuits providing reasonable approximations to the ground state energies.

Figure~\ref{figure:training-detail} provides a detailed analysis of the absolute error from FCI on a logarithmic scale for each molecule. The blue dashed horizontal line represents the chemical accuracy threshold (1.6 mHa), which is a standard benchmark for quantum chemistry calculations. The plots also include reference lines for HF and CCSD methods to provide context for the GQE performance. The detailed error analysis reveals important insights about the performance of GPT-QE across different molecular systems and bond lengths. For $\texttt{H}_2$ and $\texttt{LiH}$, chemical accuracy (1.6 mHa) is achieved across all target bond lengths, demonstrating the effectiveness of GPT-QE for these relatively simple molecular systems. However, for $\texttt{BeH}_2$ and $\texttt{N}_2$, the results show a more nuanced behavior: chemical accuracy is maintained near the equilibrium geometry, but deteriorates in the bond dissociation region where the electronic structure becomes more complex. Notably, for $\texttt{N}_2$, GPT-QE achieves higher accuracy than CCSD across most bond lengths, particularly in the equilibrium region, highlighting the potential of the generative approach to capture complex electronic correlations that are challenging for traditional quantum chemistry methods.

To further investigate the effectiveness of different loss functions in GPT-QE, we conducted a comparative study between the GRPO and Logit Matching loss functions introduced in Appendix~\ref{app:loss_functions} for the $\texttt{N}_2$ molecule at bond length 1.5 Å. Figure~\ref{fig:gate_comparison} shows the best energy relative to the ground truth energy as a function of the number of gates (circuit depth). GRPO consistently outperforms Logit Matching across all gate counts, achieving lower energy values and demonstrating better convergence properties. The chemical accuracy threshold (1.6 mHa) is indicated by the blue dashed line, with the shaded region representing the chemical accuracy regime. This comparison highlights the superior performance of GRPO loss in guiding the generative model towards more accurate quantum circuit generation across different circuit depths.

\subsection{The Effect of Pre-Training}
\label{section:pretraining}

To investigate the effectiveness of pre-training in GPT-QE, we employed the coefficient reweighting technique described in Appendix~\ref{app:coefficient_reweighting}. This approach allows us to transform 75,000 (sequence, energy) pairs obtained from 1,500 epochs of training for $\texttt{N}_2$ at bond length 1.2 Å into data suitable for bond length 1.05 Å. To retain only the most useful data, we selectively keep the top $x$\% of the lowest energy data. 

Table~\ref{tab:pretraining_data_stats} summarizes the statistics of the pre-constructed datasets used for different pre-training ratios. The datasets contain sequences with energies ranging from the minimum energy of -107.4631 Hartree to higher energy values, with the mean energy increasing as the percentage of retained data increases. Note that the FCI energy for this system is -107.5854 Hartree, indicating that the pre-constructed data consists of significantly higher energy configurations compared to the exact ground state.

\begin{table*}[h]
\centering
\caption{Statistics of pre-constructed datasets for different pre-training ratios.}
\label{tab:pretraining_data_stats}
\begin{tabular}{lcccc}
\hline
Pre-training Ratio & \# of Sequences & \multicolumn{3}{c}{Energy (Hartree)} \\
\cline{3-5}
& & Min & Mean & Std \\
\hline
Top 5\% & 3,760 & -107.4631 & -107.4193 & 0.0060 \\
Top 10\% & 7,520 & -107.4631 & -107.4165 & 0.0051 \\
\hline
\end{tabular}
\end{table*}

For incorporating the pre-constructed data into the training process, there are two possible approaches: (i) initially loading pre-constructed data and then starting to load model-generated data, or (ii) mixing a certain proportion of pre-constructed data into the model-generated data during training. We adopted approach (ii): mixing a certain proportion of pre-constructed data into the model-generated data during training. Specifically, we initially set 30\% of the replay buffer data to be pre-constructed data, with the remaining 70\% generated by the model. The proportion of pre-constructed data is then linearly decreased over 150 epochs until it reaches 0\%. This approach serves as a demonstration of pre-training capabilities, and finding the optimal strategy requires further investigation.

\begin{figure}[h]
\centering
\includegraphics[width=0.5\textwidth]{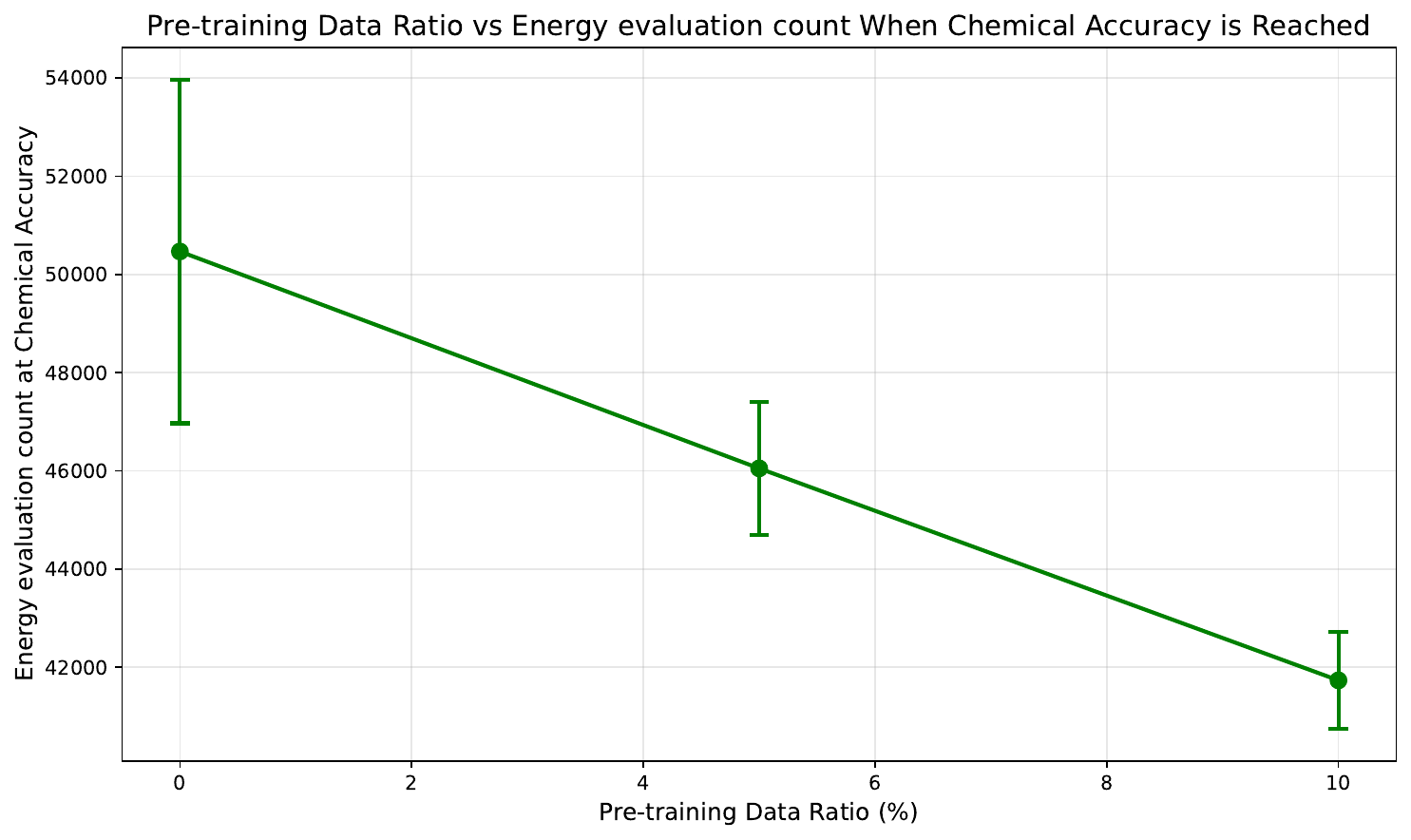}
\caption{Effect of pre-training data ratio on energy evaluation count. The plot shows the number of energy evaluations required to reach chemical accuracy as a function of the percentage of low-energy data used for pre-training. Error bars represent standard error across multiple runs.}
\label{fig:pretraining_effect}
\end{figure}

Figure~\ref{fig:pretraining_effect} shows the number of energy evaluations required to reach chemical accuracy for the $\texttt{N}_2$ molecule at bond length 1.05 Å as a function of the percentage $x$ of top low-energy data used for pre-training. The 0\% case represents training without any pre-constructed data. 

The results demonstrate that pre-constructed data significantly reduces the number of energy evaluations needed to achieve chemical accuracy. Specifically, using 5\% of the lowest energy data reduces the energy evaluation count by approximately 8\%, while using 10\% of the data reduces it by approximately 17\% compared to the baseline without pre-training. Overall, the pre-constructed data enables a reduction of nearly 20\% in the energy evaluations required to reach chemical accuracy, highlighting the effectiveness of pre-training in improving the efficiency of the GPT-QE training process.

\subsection{Demonstrations on quantum device}
To validate the practical applicability of GPT-QE, we conducted experiments on the IBM 127-qubit backend \texttt{ibm\_kawasaki} for the $\texttt{H}_2$ molecule. We selected four qubits (Qubits [1,2,3,4]) with a linear layout characterized by low readout error rates and low two-qubit error rates and to minimize the impact of device noise on our calculations. The expectation value of each Pauli term in the Hamiltonian was measured using 8,192 shots per measurement.

Given the inherent noise in current quantum devices, we applied readout error mitigation and Zero-Noise Extrapolation (ZNE) techniques \cite{Kandala2019ZNE}, as implemented in Qiskit 1.1.1 \cite{qiskit2024}, to reduce systematic errors caused by device noise. To further minimize the impact of device noise on calculation accuracy, we employed an adaptive training strategy: we initially set the number of tokens to 2 and gradually increased it until the energy values converged. For each fixed sequence length, GPT-QE was trained for 25 steps with $N_{\text{batch}}=N_{\text{sample}}=N_{\text{buffer}}=10$ and $N_{\text{iter}}=1$. The training utilized Logit Matching loss as described in Appendix~\ref{app:loss_functions}.

Figure~\ref{fig:gpt_real_device} presents the results for $\texttt{H}_2$ molecular energies across different bond lengths. The calculations performed on $ibm\_kawasaki$ closely reproduce the exact energy values, demonstrating successful training on the quantum device. We compared the quantum device results with statevector simulations using the same optimal sequences obtained from GPT-QE calculations. The GPT-QE energies from the quantum device lie within 0.002 Hartree of the ground-state energies from statevector simulations, confirming the effectiveness of our error mitigation approach.

We observed that the required sequence length increases with the $\texttt{H}_2$ bond length. Specifically, for bond lengths shorter than 0.7\AA, 2 tokens suffice, while for bond lengths of 1.2\AA~and 2.0\AA, the required number of tokens increases to 3 and 6, respectively. This trend reflects the growing importance of electron correlation effects, which Hartree-Fock theory cannot capture. GPT-QE effectively captures these correlation effects by adaptively increasing the number of tokens, demonstrating its ability to handle varying levels of electronic complexity.

\begin{figure}[h]
\centering
\includegraphics[width=0.5\textwidth]{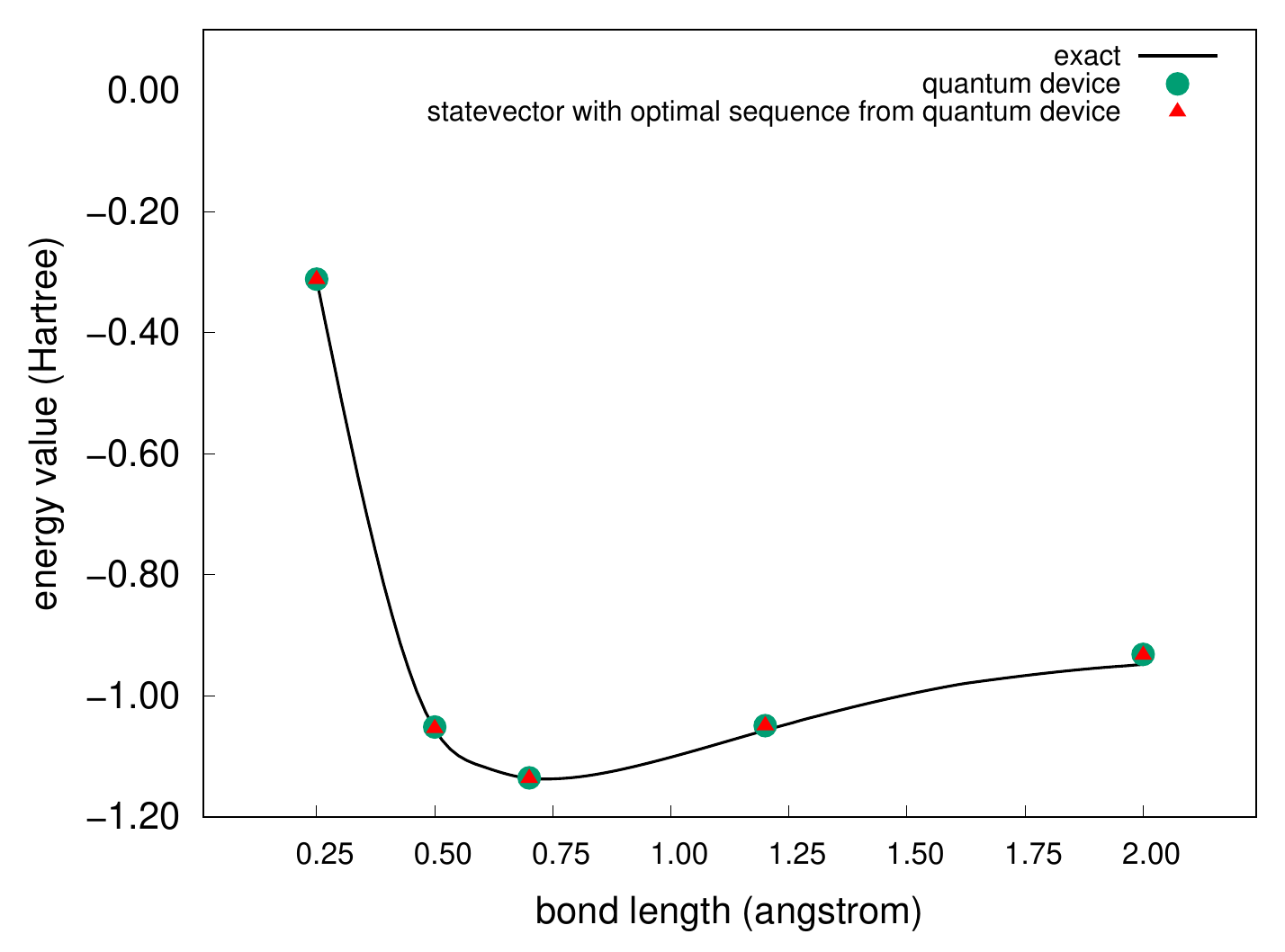}
\caption{Comparison of $\texttt{H}_2$ molecular energies at different bond lengths: exact diagonalization results (black line), GPT-QE calculations on the IBM quantum device with error mitigation (green spheres), and statevector simulations using optimal sequences from quantum device training (red triangles). The quantum device results closely match the exact energies, demonstrating successful training and error mitigation.}
\label{fig:gpt_real_device}
\end{figure}

\section{Conclusion and Discussion}
\label{section:conclusion}
We propose the GQE algorithm, a novel method that applies a generative model to obtain quantum circuits with desired properties. In particular, we introduce GPT-QE, which is based on the transformer architecture, and we design its training scheme. We also introduce and discuss proposals for extending the architecture's capabilities by including inputs. In our numerical experiments, we address the problem of searching for the ground state of the electronic structure Hamiltonians of several molecules: $\texttt{H}_2$, $\texttt{LiH}$, $\texttt{BeH}_2$, and $\texttt{N}_2$ . We demonstrate that GPT-QE finds a quantum state with energy close to that of the ground state.

\label{section:other_research}
Many other research directions should be explored to fully enable the GPT-QE models to progress beyond this proof of concept. First of all, conditional generalization described in Section~\ref{section:transfer} is an important direction to fully utilize the power of machine learning.
Additionally, experiments with larger molecules are required to further assess the optimization behavior. Another consideration is how to effectively integrate GPT-QE with the VQE framework. A straightforward approach might involve using VQE as a post-processing step, but there are many opportunities for more integrated hybridization. As a particularly interesting existing method for hybridization, the ADAPT-VQE strategy is known to achieve high accuracy \cite{grimsley2019adaptive}, though it necessitates numerous quantum circuit runs. Combining ADAPT-VQE with GPT-QE is another potential direction for future research.

Each component of GPT-QE is also open to updates and improvements. In our numerical experiment, we employ a chemically inspired operator pool, but, theoretically, any kind of operators could be included. The transformer's design could be modified to generate a sequence of tokens and the parameters embedded in quantum gates. Such an extension would facilitate easier hybridization with VQE. To fully leverage the transformer's capabilities, exploring how to design suitable inputs for the transformer is also crucial.

As mentioned earlier in the paper, applying and extending the GQE framework to problems beyond ground-state approximation is also feasible. For instance, Ref. \cite{minami2025generative} has already investigated the application of a GQE model with classical inputs to the problem of binary optimization. One could also envision applying such input-conditioned GQE models to supervised machine-learning problems. The critical question would then be determining which types of machine learning problems are best suited for the GQE framework. This inquiry requires careful study and identification of suitable problems.

We hope that, in a parallel track, and a decade later from when some of us introduced VQE, the community will embrace GQE and work together to enable many of the extensions briefly proposed above to help in achieving the goal of near-term quantum utility.

\section*{Acknowledgements}
This research used resources of the National Energy Research
Scientific Computing Center, a DOE Office of Science User Facility
supported by the Office of Science of the U.S. Department of Energy
under Contract No. DE-AC02-05CH11231 using NERSC award
NERSC DDR-ERCAP0027330.  K.N. acknowledges the support of Grant-in-Aid for JSPS Research Fellow 22J01501. L.B.K. acknowledges support from the Carlsberg Foundation. A.A.-G.
acknowledges support from the Canada 150 Research Chairs program and CIFAR as well as the generous support of Anders G. Fr\o seth.
We are deeply grateful to the Defense Advanced Research Projects Agency (DARPA) for their generous support and funding of this project, under the grant number HR0011-23-3-0020. A part of this work was performed for Council for Science, Technology and Innovation (CSTI), Cross-ministerial Strategic Innovation Promotion Program (SIP), “Promoting the application of advanced quantum technology platforms to social issues” (Funding agency: QST).

We acknowledge that nearly simultaneously with the second version update of this work, \cite{nakamura2025persistent} proposed improvements to our original GQE learning approach by incorporating mechanisms equivalent to modified loss functions and replay buffers.

\bibliographystyle{ieeetr}
\bibliography{main}

\appendix
\onecolumn

\section{Details of the experiment}
\label{app:experiment_details}
\subsection{Molecule Generation}
\label{app:molecule_generation}

The molecular systems are generated using the CUDA Quantum solver \cite{cuda-quantum} with the following configuration:

\begin{lstlisting}[language=Python]
import cudaq_solvers as solvers

molecule = solvers.create_molecule(geometry, molecule.basis, spin=0, charge=0, nele_cas=nele_cas, norb_cas=norb_cas, ccsd=True, casci=True)
\end{lstlisting}
where \texttt{geometry} is the geometry of the molecule, \texttt{molecule.basis} is the basis set, \texttt{nele\_cas} is the number of electrons in the active space, and \texttt{norb\_cas} is the number of orbitals in the active space.
The molecule object generation includes the calculation of the Hartree-Fock energy, the CCSD energy, and the FCI energy based on PySCF \cite{sun2020recent}.

\subsection{Operator Pool Configuration}
\label{app:operator_pool}

Our operator pool consists of Pauli time evolutions plus the identity operator:
$\mathcal{G} = \{e^{i \hat{P}_j t_j} \}_{j} \cup \{I\}$, 
where $\hat{P}_j$ represents a tensor product of Pauli operators, $t_j$ is a real number, and $I$ is the identity operator. Defining $\mathcal{P}$ as the set of $\hat{P}_j$ and $\mathcal{T}$ as the set of $t_j$, the size of the operator pool is $|\mathcal{P}| \times |\mathcal{T}| + 1$. 

For $\mathcal{P}$, we derive chemically inspired choices from the unitary coupled-cluster single and double excitations (UCCSD). Letting $T$ denote the sum of all fermionic excitation operators included in UCCSD, $\mathcal{P}$ is selected such that $e^{i P_\ell \theta_\ell} (P_\ell \in \mathcal{P})$, with an angle $\theta_\ell$, is part of the decomposed operators when $e^{T - T^{\dagger}}$ is broken down into Pauli time evolutions by the Trotter decomposition \cite{trotter1959product}. For $\mathcal{T}$, we choose $\mathcal{T} = \left\{\pm 2^k/320 \right\}_{k=0}^5$. In our implementation, we use the CUDA Quantum solver \cite{cuda-quantum} to generate the operator pool:
\begin{lstlisting}[language=Python]
import cudaq_solvers as solvers
import numpy as np

pool = []
params = [2**k/320 for k in range(0, 5)] + [-2**k/320 for k in range(0, 5)]
operators = solvers.get_operator_pool("uccsd", n_qubits, n_electrons)
pool.append(get_identity(n_qubits))
for o in operators:
	for p in params:
		pool.append(exp_pauli(p * o)) # Create a scaled Pauli operator
\end{lstlisting}
where \texttt{n\_qubits} is the number of qubits in the quantum system and \texttt{n\_electrons} is the number of electrons in the molecule. Here, \texttt{get\_identity(n\_qubits)} returns the identity operator for the quantum system, and \texttt{exp\_pauli(P)} implements the exponential operator $e^{iA}$ for a given 
operator $A$.

\subsection{Adaptive scheduling of inverse temperature $\beta$}
\label{app:beta_adaptive}

In the GPT-QE setup, the inverse temperature $\beta$ governs the exploration-exploitation trade-off. We employ a dispersion-triggered schedule that adapts $\beta$ based on the variance of the energies generated at each training iteration. Let $\{E(\vec{j}_m)\}_{m=1}^{M}$ be the energies evaluated at iteration $t$ and define
\begin{equation}
\label{eq:beta_adaptive}
\beta_{t+1}=
\begin{cases}
\beta_t-\alpha, & \text{if } \operatorname{std}\!\left(\{E(\vec{j}_m)\}_{m=1}^{M}\right) < \tau_{\mathrm{disp}},\\[2pt]
\beta_t+\alpha, & \text{otherwise},
\end{cases}
\end{equation}
with a fixed step size $\alpha=0.02$ and dispersion threshold $\tau_{\mathrm{disp}}=10^{-5}$ in all experiments. Intuitively, when the batch collapses (very small dispersion), we decrease $\beta$ to re-broaden the sampling distribution and preserve exploration; otherwise, we increase $\beta$ to gradually sharpen the distribution and intensify exploitation around lower-energy regions discovered so far. Compared to a monotonic temperature-increase schedule, this dispersion-triggered update is less prone to entrapment in local minima and empirically improves convergence. Note that this strategy utilizes the fact that the model only needs to sample the optimal circuit a high fraction of the time, but does not need to converge to a collapsed distribution sampling \emph{only} the optimal circuit.

\section{Loss Functions}
\label{app:loss_functions}

\subsection{Logit Matching Loss}
\label{app:logit_matching}

The logit-matching loss is designed to align the model's cumulative logits $w_{\mathrm{sum}}(\vec{j}_m;\theta)$ with the evaluated energies $E(\vec{j}_m)$, inducing a pseudo-thermal preference for lower-energy circuits. This approach encourages the model to generate circuits with lower energies by making the logits correspond to the energy landscape.

Logit matching is inspired by the following argument. Let us consider the process sampling $U_N(\vec{j})$ according to $p_N(\beta, \vec{j})$ and applying it to $\rho_0$. Let $\rho(\beta)$ be the quantum state generated by the stochastic process. With $\mathcal{E}_N(\vec{j}, \rho) := U_N(\vec{j}) \rho U_N(\vec{j})^{\dagger}$, it can be written as 
\begin{equation}
\begin{split}	
\rho(\beta) &= \sum_{j} p_N(\beta, \vec{j})\mathcal{E}_N(\vec{j}, \rho_0) \\
&=  \frac{1}{\mathcal{Z}}\sum_{j} \exp\left(-\beta w_{\textrm{sum}}(\vec{j})\right)
\mathcal{E}_N(\vec{j}, \rho_0).
\end{split}
\end{equation}
We observe that if $w_{\textrm{sum}}(\vec{j}) = E(\vec{j})$ is satisfied, $\rho(\beta)$ gives a pseudo thermal state with the inverse temperature $\beta$ in the sense that the quantum state is generated according to the probability $\exp\left(-\beta E(\vec{j})\right)$. Therefore, increasing the value of $\beta$ creates a bias towards generating lower energy quantum states.

From these observations, we design the logit matching loss so that $w_{\textrm{sum}}(\vec{j}) \simeq E(\vec{j})$ is satisfied. The loss function is defined as
\begin{equation}
\label{eq:costTrainingApp}
\mathcal{L}_{\textrm{LM}} = 
\frac{1}{N_{\textrm{batch}}} \sum_{m=1}^{N_{\textrm{batch}}} \left(e^{-\beta w_{\textrm{sum}}(\vec{j}_m)} - e^{-\beta E(\vec{j}_m)} \right)^2.
\end{equation} 

For numerical stability, we add an offset to the output of the quantum device. More specifically, when calculating the cost function above, we substitute $E(\vec{j}) + E_{\textrm{offset}}$ instead of the original $E(\vec{j})$. The value of $E_{\textrm{offset}}$ is chosen to be 107 for $\texttt{N}_2$, in the experiment for Fig.~\ref{fig:gate_comparison}. Since we match the sum of logits with the energy function, we call this technique \textit{logit-matching}.

\subsection{GRPO-based Loss Function}
\label{app:pg_guidance}

While the logit-matching loss (Eq.~\eqref{eq:costTrainingApp}) encourages a pseudo-thermal preference, it does not explicitly minimize the expected energy; empirically, this can lead to periods where the loss decreases yet the energy plateaus above the ground-state level. To remedy this, an energy-oriented policy surrogate is added to amplify the probability of lower-energy samples within each iteration. Specifically, we instantiate this surrogate using the Group Relative Policy Optimization (GRPO) \cite{GRPO2024} objective, which computes so-called advantages by normalizing rewards within a group of sampled sequences. For a sequence $\vec{j}_m=\{j_{m,k}\}_{k=1}^N$, define the sequence-level reward and advantage
\begin{equation}
\label{eq:reward_adv} % (10)
r_m := -E(\vec{j}_m),\qquad
\hat{A}_m:=\frac{r_m-\operatorname{mean}(\{r_{m}\}_{m=1}^{M})}{\operatorname{std}(\{r_{m}\}_{m=1}^{M})},
\end{equation}
which yields variance reduction and scale invariance across iterations. We further define the importance ratio
\begin{equation}
\label{eq:importance_ratio}
\rho_{m,k}\;:=\;
\frac{\pi_{\theta}(j_{m,k}\mid q, j_{m,<k})}{\pi_{\theta_{\mathrm{old}}}(j_{m,k}\mid q, j_{m,<k})}.
\end{equation}
The function
$\pi_\theta (j_{m,k}\mid q, j_{m,<k})$
is the probability function that $j_{m,k}$ is generated as the $k$-th token given the $k-1$ previously generated tokens $j_{m,<k}$ and a fixed start token, and it corresponds to $\exp\left(-\beta w^{(k)}_{j_{m,k}}\right)$ in the notation introduced in Section~\ref{section:gptqe-sampling}. The function has the parameters $\theta$ included in the neural network written out explicitly. 
The symbol $\theta_{\mathrm{old}}$ denotes frozen reference parameters, taken as the initial parameters in all of our experiments. Using $\rho_{m,k}$, the individual GRPO-based loss function for each batch is defined as
\begin{equation}
\label{eq:grpo}
\mathcal{L}_{\text{GRPO}}= \frac{1}{N_{\textrm{batch}}}\sum_{m=1}^{N_{\textrm{batch}}} \frac{1}{N}
\sum_{k=1}^{N}
\mathrm{clip}\left(
\rho_{m,k},\,1-\varepsilon,\,1+\varepsilon
\right)\, \hat{A}_{m},
\end{equation}
where $\textrm{clip}$ denotes the clipping operation, which restricts the value of 
$\rho_{m,k}$ to lie within the interval $[1-\varepsilon,\,1+\varepsilon]$ with $\varepsilon\in(0,1)$ the clipping parameter (we use $\varepsilon=0.2$). 
Specifically, it returns $\rho_{m,k}$ if $1-\varepsilon \leq \rho_{m,k} \leq 1+\varepsilon$, 
the lower bound $1-\varepsilon$ if $\rho_{m,k} < 1-\varepsilon$, 
and the upper bound $1+\varepsilon$ if $\rho_{m,k} > 1+\varepsilon$.

\section{Coefficient reweight to Increase the Dataset}
\label{app:coefficient_reweighting}
In the context of quantum computation, the electronic structure Hamiltonian can be mapped to a linear combination of Pauli operators:
\begin{equation}
\hat{H}(\vec{\Delta}) = \sum_{a=1}^{N_H} h_a(\vec{\Delta}) \hat{P}_a,
\end{equation}
where $h_a(\vec{\Delta})$ are real coefficients when the geometry is $\vec{\Delta}$ and $\hat{P}_a$ are tensor products of Pauli operators. When we evaluate the expectation value,
we need to measure each individual Pauli expectation value $\langle \hat{P}_a \rangle$. Therefore, in the training of GPT-QE with $\hat{H}(\vec{\Delta})$, we obtain the dataset $\{\vec{j}_m , q_a(\vec{j}_m)\}_{m,a}$ as well as $\{\vec{j}_m, E(\vec{\Delta}, \vec{j}_m) \}_m$, where $q_a(\vec{j}_m)$ is the estimated value of $\langle \hat{P}_a\rangle$ when the circuit is constructed with the sequence of tokens $\vec{j}_m$ and
$E(\vec{\Delta}, \vec{j}_m) = \sum_a h_a(\vec{\Delta}) q_a(\vec{j}_m)$.

The estimated values ${q_a (\vec{j_m})}_{m,a}$ can be used to construct $\{\vec{j}_m, E(\vec{\Delta}^{\prime}, \vec{j}_m)\}_{m}$ with a different geometry $\Delta^{\prime}$ by simply combining the measured Pauli expectation values using different coefficients $h_a(\vec{\Delta}^{\prime})$.
It can then be used for pre-training when solving the ground-state search of the Hamiltonian $H(\vec{\Delta}^{\prime})$.
\end{document}